# Dosimetric Evaluation of CT-to-RED Calibration Curves on Dosimetric Accuracy in Brain Radiotherapy

Hossam Donya[1, 2*], Duong Thanh Tai[3], Islam G. Ali [4]

1 2 3 4 Affiliations were removed for blinded review in the journal

**Abstract**

**Background:** Accurate dose calculation in radiotherapy relies on converting Hounsfield Units (HU) from CT images to relative electron density (RED) through calibration curves. However, the selection of calibration methods, particularly the use of phantom-derived versus default TPS curves, can significantly impact dosimetric accuracy, especially in complex anatomical regions like the brain.

**Purpose**: To evaluate the dosimetric impact of different CT-to-RED calibration curves—derived from the CIRS 062M phantom, the Catphan 604 phantom, and the default Monaco treatment planning system (TPS) curve—on dose calculation in clinical brain radiotherapy plans.

Methods: Ten volumetric modulated arc therapy (VMAT) plans for brain tumors were retrospectively recalculated using three CT-to-RED calibration curves. Dose-volume histogram (DVH) metrics were extracted for the planning target volume (PTV) and ten critical organs at risk (OARs). Pairwise statistical comparisons were conducted to assess differences in Dmax, Dmin, Dmean, and D95% using repeated-measures ANOVA and Wilcoxon signed-rank tests, with $p < 0.05$ considered significant.

**Results:** Significant differences were found in all PTV dose metrics between the CIRS and default calibration curves ($p < 0.05$), with the default curve overestimating doses by 3–5%. Similarly, Dmax differences of 2–4% were observed in multiple OARs, including the brainstem ($p = 0.003$) and optic chiasm ($p = 0.005$). The Catphan-derived curve showed intermediate agreement, with fewer statistically significant differences.

**Conclusion:** The choice of CT-to-RED calibration curve substantially affects dose calculation accuracy in brain radiotherapy. The CIRS-derived curve provided the most accurate dosimetric estimates and should be prioritized in clinical calibration protocols. Adopting site-specific calibration methods may reduce inter-institutional variability and improve treatment safety in precision radiotherapy.

**Keywords:** Brain radiotherapy; CT-to-RED calibration; Electron density; Treatment planning system; Dosimetric accuracy; CIRS phantom; Catphan phantom; VMAT; Organs at risk (OARs)

## 1. Introduction

Accurate dose calculation is fundamental to effective radiotherapy, ensuring optimal tumor control while sparing adjacent healthy tissues as much as possible [1]. In brain radiotherapy, where intricate anatomy and critical structures such as the brainstem, optic nerves, and lenses are in close proximity, dosimetric precision is paramount. International guidelines stipulate that delivered radiation doses should not deviate by more than 5% from the prescribed dose, as even minor inaccuracies can reduce tumor control probability by approximately 20% [2]. Central to this precision is the conversion of Hounsfield Units (HU) from computed tomography (CT) scans to relative electron density (RED), a critical step in modeling radiation interactions within treatment planning systems (TPS) [3].

CT imaging provides essential anatomical and quantitative data for radiotherapy planning [4]. This challenge is further exacerbated in brain radiotherapy, where the coexistence of compact bone, cerebrospinal fluid, and neural tissues within a confined volume intensifies the sensitivity of dose calculations to HU-to-RED conversion inaccuracies. HU values, which reflect X-ray attenuation properties of tissues, must be converted to RED to account for the energy-dependent interactions of therapeutic megavoltage radiation beams, primarily governed by Compton scattering, as opposed to the photoelectric effect dominant in diagnostic CT [5]. RED, a dimensionless quantity representing the electron density of tissue relative to water, directly influences photon scattering and absorption, enabling accurate dose calculations [6]. CT-to-RED calibration curves, derived from phantoms or provided as default by TPS manufacturers, facilitate this conversion. However, inaccuracies in these curves can lead to significant dosimetric errors, particularly in complex regions like the brain [7,8]. CT-to-RED calibration curves can either be generated from direct phantom measurements or adopted from default TPS presets. Phantom-based calibrations are typically constructed using tissue-equivalent materials to closely approximate real patient heterogeneities. In contrast, default calibration curves, while convenient, are often based on generic population data and may not accurately represent specific anatomical sites such as the brain.

Numerous studies have explored CT-to-RED calibration methodologies. Annkah et al. (2014) compared the Catphan 504 and CIRS 062 phantoms for kV-cone beam CT (CBCT) dose calculations, reporting dose discrepancies of up to ±14% with Catphan-derived curves versus ±5% with CIRS-derived curves compared to CT-based planning [9]. These findings highlight the importance of phantom selection for accurate calibration, especially in high-precision sites like the brain. Zurl et al. (2014) demonstrated that HU variations significantly affect CT-to-RED conversion and dose distributions [10]. Additionally, dual-energy CT (DECT) has shown promise in improving electron density estimation, particularly in high-density tissues such as bone, as reported by Wohlfahrt et al. (2017) [11]. A recent study by Duong et al. (2025) [12] examined the influence of CT scanning protocols—specifically tube voltage and current—on HU-to-RED calibration curves and radiotherapy dose calculation for the Halcyon LINAC system. Their phantom-based results confirmed that variations in scanning parameters, particularly at low tube voltages, significantly affected RED values in high-density tissues, although the impact on overall dose calculation was minimal (within 1.37%). These studies underscore the need for site-specific calibration to optimize dosimetric accuracy. Furthermore, inconsistent calibration practices across institutions may hinder reproducibility in multi-center clinical trials and benchmarking studies, further highlighting the need for standardized, validated calibration protocols.

Despite these advances, a critical knowledge gap persists. Most prior studies have been phantom-based or focused on CBCT, with limited evaluation of CT-to-RED calibration curves in clinical volumetric modulated arc therapy (VMAT) plans for brain radiotherapy. Default TPS calibration curves may not adequately represent the brain's complex tissue composition, including bone and dense soft tissues, potentially compromising treatment outcomes [13,14]. Furthermore, the comparative dosimetric impact of phantom-derived versus default calibration curves in clinical brain VMAT plans remains underexplored.

This study aims to address this gap by evaluating the impact of three CT-to-RED calibration curves—derived from the CIRS Model 062M Electron Density Phantom, the Catphan 604 Phantom, and the default curve in the Monaco TPS—on dose calculation accuracy in clinical brain

VMAT treatment plans. By analyzing dosimetric outcomes for the Planning Target Volume (PTV) and critical Organs at Risk (OARs) in ten brain cancer cases, we seek to determine whether phantom-specific calibration introduces clinically significant deviations, assess the extent to which calibration strategies affect dose accuracy, and evaluate the necessity of site-specific calibration in modern TPS. These findings aim to inform calibration practices, enhancing the safety and efficacy of brain radiotherapy.

## 2. Materials and Methods

### *2.1. CT Image Acquisition and Phantom Scanning*

CT images were acquired using a Siemens Somatom CT scanner (Siemens HealthCare, Erlangen, Germany). All scans were performed with a slice thickness of 2 mm, 120 kVp tube voltage, and standard soft-tissue reconstruction algorithms. Two phantoms were used to generate CT-to-RED calibration curves: the Catphan 604 (The Phantom Laboratory, USA) and the CIRS Model 062M Electron Density Phantom (CIRS, USA)

### *2.2. CT-to-RED Calibration Curve Generation*

The Catphan phantom includes seven inserts with distinct attenuation properties (air, PMP, LDPE, polystyrene, acrylic, Delrin, and Teflon), while the CIRS phantom contains ten tissue-equivalent materials with known RED values (ranging from lung to dense bone). Calibration curves were constructed by plotting mean HU values versus corresponding REDs, obtained from circular ROIs drawn on CT images. The default HU-to-RED calibration curve from Monaco TPS version 6.1.2 was also extracted for comparison.

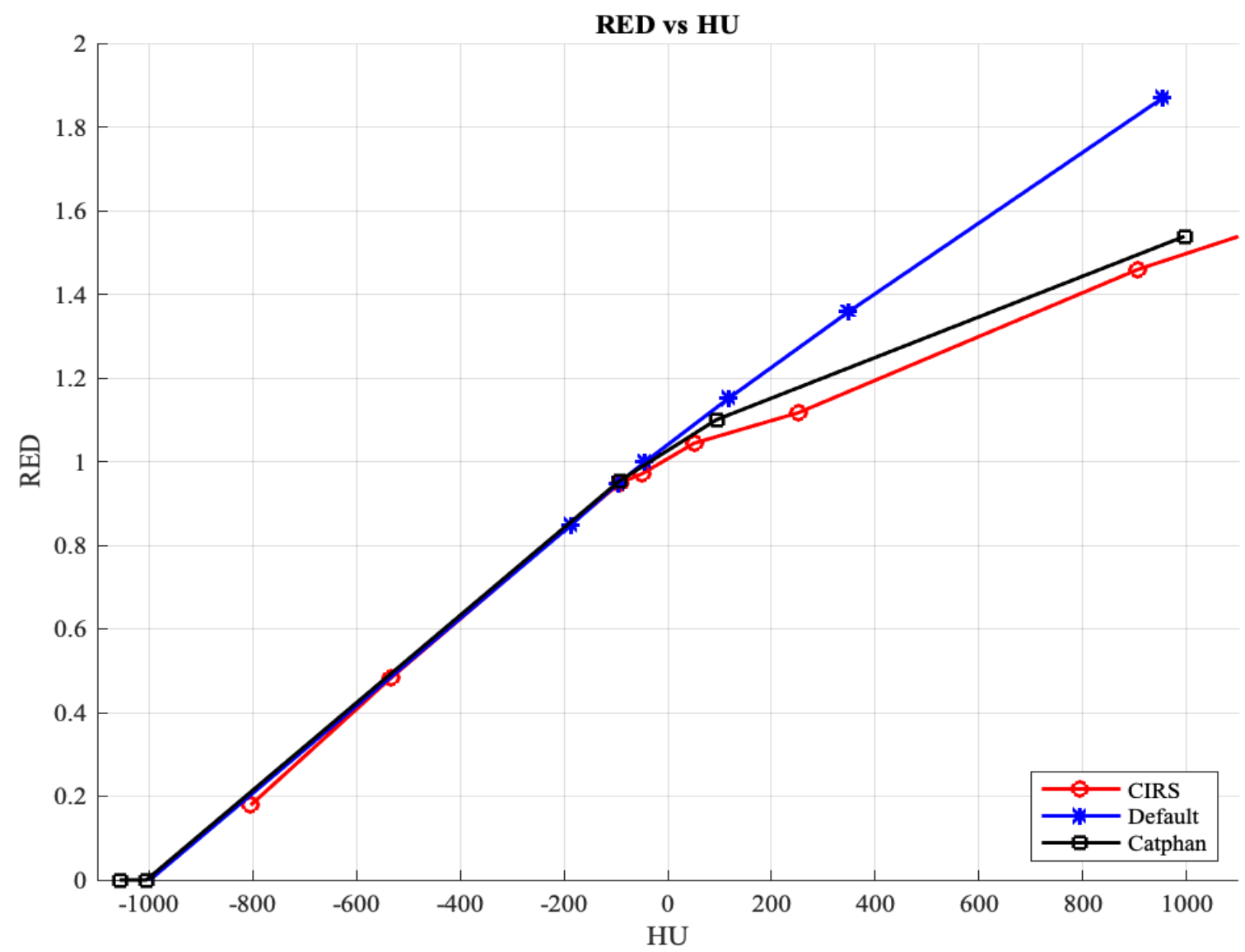

Figure 1. Comparison of CT-to-RED calibration curves derived from the CIRS phantom (red), Catphan phantom (black), and the default Monaco TPS curve (blue). The RED values were obtained by mapping the average Hounsfield Units (HU) of each phantom insert to their corresponding relative electron densities.

*2.3. Patient Selection and Treatment Planning*

Ten brain tumor patients previously treated with VMAT were retrospectively enrolled in this study cohort. Treatment planning was performed using the Monaco™ V6.1.2 Treatment Planning System, applying the Monte Carlo dose calculation algorithm. All plans were created using a 6 MV photon beam delivered by the Elekta Synergy Agility LINAC (160 MLC leaves). Each plan was recalculated using the three CT-to-RED calibration curves (Catphan, CIRS, default) while keeping beam parameters and monitor units constant.

*2.4. Dosimetric Data Extraction*

Dose-volume histogram (DVH) data were extracted for each treatment plan to evaluate the dosimetric impact of different CT-to-RED calibration curves. For the planning target volume (PTV), the following parameters were analyzed: maximum dose (Dmax), minimum dose (Dmin), mean dose (Dmean), and the dose received by 95% of the volume (D95%). To assess the potential impact on surrounding normal tissues, the maximum dose (Dmax) was also recorded for critical organs at risk (OARs), including the brainstem, optic nerves, optic chiasm, cochleas, eyes, and lenses. All dosimetric evaluations and reporting adhered to the recommendations outlined in ICRU Report 83, ensuring consistency and clinical relevance in interpreting the results.

### *2.5. Statistical Analysis*

All statistical analyses were performed using IBM SPSS Statistics version 30.0.0.0 (IBM Corp., Armonk, NY, USA). The Shapiro–Wilk test was applied to assess the normality of the dosimetric data distributions. For datasets that followed a normal distribution, repeated-measures analysis of variance (ANOVA) was conducted, followed by Bonferroni-adjusted paired t-tests for post-hoc comparisons. In cases where the data violated the assumption of normality, the non-parametric Friedman test was employed, and significant differences between groups were further examined using Wilcoxon signed-rank tests. A two-sided p-value of less than 0.05 was considered indicative of statistical significance.

## **3. Results**

This study evaluated the dosimetric effects of three distinct CT-to-RED calibration curves: those derived from the CIRS 062M phantom, the Catphan 604 phantom, and the default calibration embedded in the Monaco Treatment Planning System (TPS). All recalculations were performed on ten clinical volumetric modulated arc therapy (VMAT) plans for brain tumors. Comparisons were based on dose-volume histogram (DVH) metrics for the planning target volume (PTV) and organs at risk (OARs).

### *3.1. PTV Dosimetric Comparisons*

All PTV dose metrics—Dmax, Dmin, Dmean, and D95%—exhibited statistically significant differences between the CIRS-derived and default TPS calibration curves ($p < 0.05$). The default calibration consistently produced higher dose values, with the greatest discrepancy observed in Dmax ($p < 0.001$). This suggests a systematic overestimation of target dose when relying on the default curve. In contrast, comparisons between the Catphan and default curves, as well as between Catphan and CIRS, yielded no statistically significant differences in most metrics, except for Dmax and D95%, which showed borderline significance (Table 1).

### *3.2. OAR Dose Variations*

Among the ten analyzed OARs, seven showed statistically significant differences in Dmax when comparing the CIRS and default calibration curves. These included the brainstem ($p = 0.003$), optic chiasm ($p = 0.005$), both cochleas, both optic nerves, and one lens. The default curve resulted in consistently higher Dmax values across these structures. The Catphan-derived curve produced intermediate values and fewer significant differences compared to CIRS. Notably, comparisons between Catphan and default curves demonstrated minimal statistically significant variation, further supporting their closer alignment.

### *3.3. Pairwise Statistical Summary*

Table 1 summarizes the p-values from all pairwise comparisons for PTV and OAR dose metrics. Statistically significant values ($p < 0.05$) are denoted in bold and italic. The CIRS curve resulted in the lowest overall dose estimates, the default curve the highest, and the Catphan curve displayed an intermediate behavior with variable agreement depending on the structure.

**Table 1.** Pairwise comparison of dose metrics (Dmax, Dmin, Dmean, D95%) for the planning target volume (PTV) and organs at risk (OARs) using calibration curves derived from CIRS, Catphan, and the default Monaco TPS. Bold and italic values indicate statistically significant differences ($p < 0.05$). Post-hoc analysis applied Bonferroni correction for parametric data and Wilcoxon signed-rank test for non-parametric data.

| | | | Pairwise Comparison Test [p-value] * | | |
|---|---|---|---|---|---|
| | Dosimetric Parameter (%) | Sigificance [p-value] | Catphan - CIRS | Catphan - Default | Default - CIRS |
| Brainstem | $D_{max}$ | ***<0.001*** | 0.061[a] | ***0.001***[a] | ***0.003***[a] |
| Chiasm | $D_{max}$ | ***<0.001*** | 0.074[b] | ***0.037***[b] | ***0.005***[b] |
| Rt. Cochlea | $D_{max}$ | ***0.0057*** | 0.237[a] | 0.318[a] | ***0.008***[a] |
| Lt. Cochlea | Dmax | ***0.0055*** | 0.241[b] | ***0.013***[b] | ***0.013***[b] |
| Rt. Nerve | $D_{max}$ | ***<0.001*** | 0.279[a] | ***0.003***[a] | ***0.021***[a] |
| Lt. Nerve | $D_{max}$ | ***<0.001*** | 0.564[a] | ***0.001***[a] | ***0.014***[a] |
| Rt. Eye | $D_{max}$ | ***0.001*** | 0.476[a] | ***0.004***[a] | ***0.013***[a] |
| Lt. Eye | $D_{max}$ | ***<0.001*** | ***0.036***[a] | 0.476[a] | ***0.004***[a] |
| Rt. Lens | $D_{max}$ | ***0.004*** | 0.108[a] | 0.682[a] | ***0.03***[a] |
| Lt. Lens | $D_{max}$ | ***0.005*** | 1[a] | ***0.022***[a] | ***0.018***[a] |
| PTV | $D_{max}$ | ***<0.001*** | ***0.035***[a] | ***0.046***[a] | ***<0.001***[a] |
| | $D_{Min}$ | ***<0.001*** | ***0.005***[b] | ***0.005***[b] | ***0.005***[b] |
| | $D_{Mean}$ | ***0.0035*** | 1[a] | ***0.001***[a] | ***0.041***[a] |
| | D95% | ***<0.001*** | ***0.015***[a] | ***0.013***[a] | ***0.006***[a] |

[a] Post-hoc pairwise comparisons test with Bonferroni correction

[b] Wilcoxon Signed Ranks Test

*P-values were derived from pairwise comparison tests conducted using both parametric repeated-measures ANOVA and non-parametric Friedman tests. Statistically significant differences are indicated with bold and italic formatting.

### *3.4. Visual Representation of Dosimetric Differences*

### 3.4. Visual Representation of Dosimetric Impact

Figure 2 presents the mean dose values and standard deviations for the PTV and OARs across the three calibration curves. These plots clearly demonstrate the dose elevation associated with the default curve, particularly for the brainstem and optic pathway structures. Figure 3 displays representative DVHs for a single patient, highlighting the deviation in dose distributions as a function of calibration method. Collectively, these findings underscore the impact of calibration curve selection on dose calculation accuracy in brain radiotherapy. The results support the use of CIRS-derived calibration for improved dosimetric precision, especially in high-gradient dose regions adjacent to critical structures.

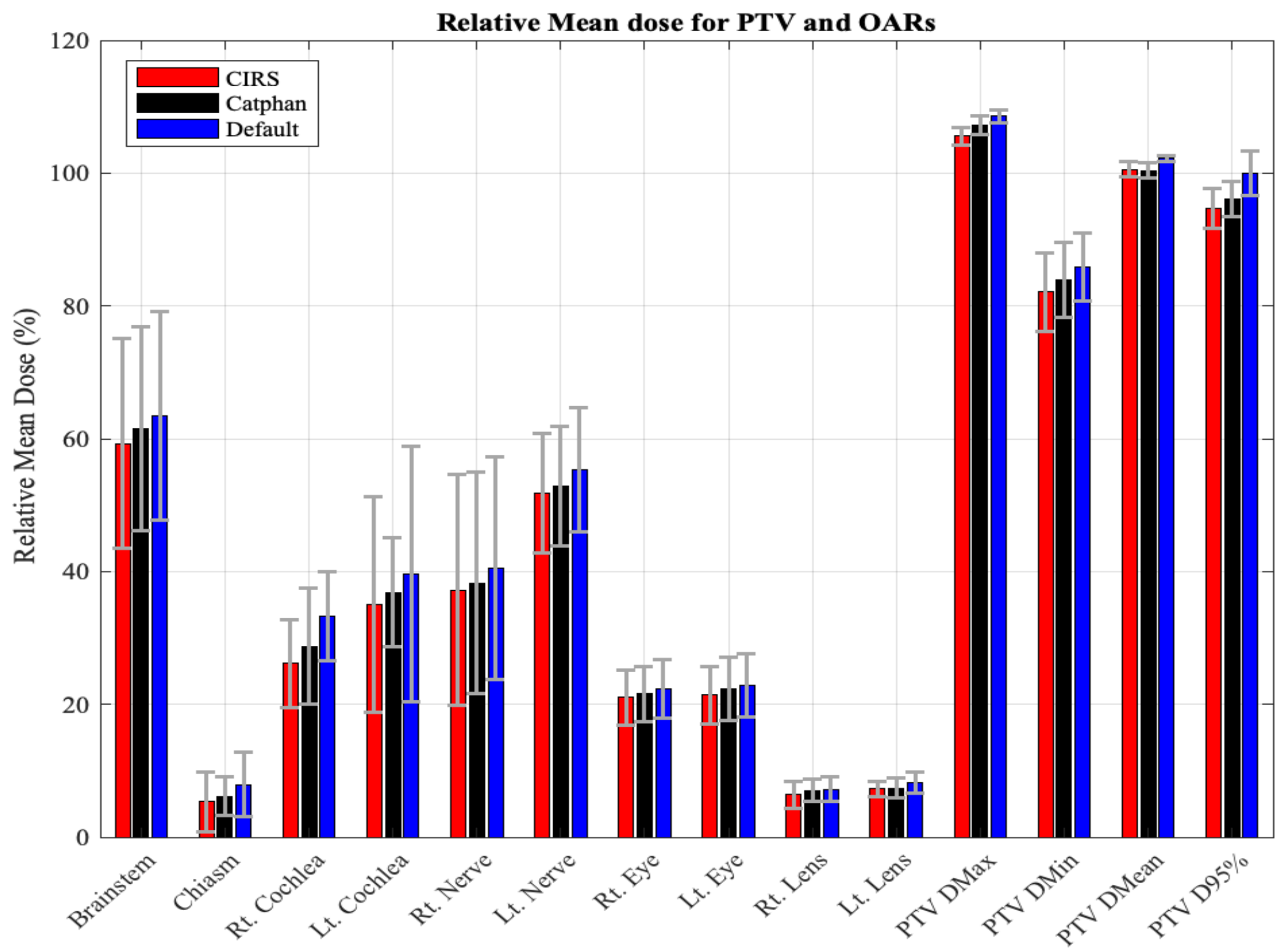

Figure 2: Grouped bar plot illustrating the comparison of relative mean dose percent for PTV ($D_{max}$, $D_{min}$, $D_{mean}$, and D95%) and the maximum doses for each OAR across the CIRS, Catphan, and Default datasets. The height of each bar represents the relative mean dose for the respective dataset, while the error bars denote the standard deviation (SD), highlighting the variability within the measurements red for CIRS, black for Catphan, and blue for Default.

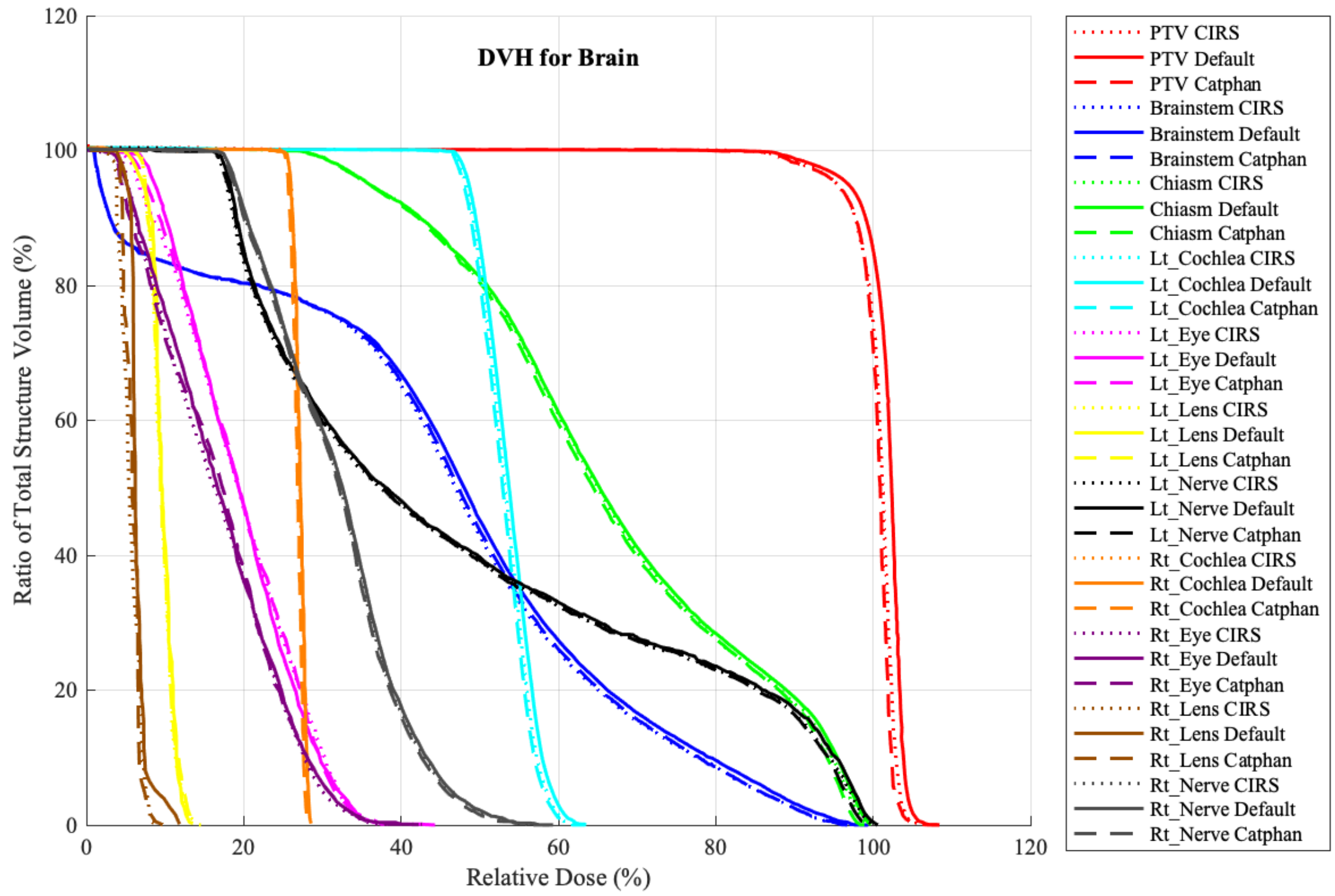

Figure 3: Comparative dose-volume histograms (DVH) for a brain case, showcasing calculations using the CIRS, Catphan, and default CT-to-RED calibration curves.

## 4. Discussion

This study underscores the critical role of site-specific CT-to-RED calibration curves in ensuring dosimetric accuracy for brain radiotherapy. The CIRS-derived calibration curve significantly outperformed the default Monaco Treatment Planning System (TPS) curve, with statistically significant differences in Planning Target Volume (PTV) and Organs at Risk (OAR) dose metrics ($p < 0.05$). These findings align with the study's objective to evaluate the dosimetric impact of calibration curves and emphasize the need for tailored calibration to optimize treatment outcomes.

The CIRS-derived curve more accurately represented brain tissue electron density, particularly in high-density regions such as bone and dense soft tissues. This is attributable to the CIRS Model 062M phantom's inclusion of ten tissue-equivalent materials, covering a broader density range than the Catphan 604 phantom's seven inserts [14]. Significant differences in PTV metrics (Dmax,

Dmin, Dmean, D95%), with Dmax discrepancies of 3–5%, and OAR Dmax values (e.g., brainstem, $p = 0.003$; optic chiasm, $p = 0.005$) between CIRS and default curves indicate that the default curve systematically overestimates doses. For instance, the default curve increased brainstem Dmax by 2–3%, potentially elevating the risk of neurotoxicity, such as brainstem necrosis. These findings are consistent with Gonod et al. (2017), who demonstrated superior dosimetric accuracy with CIRS phantoms in brain radiotherapy [15].

The Catphan-derived curve exhibited closer agreement with the CIRS curve than the default curve, with fewer significant differences in PTV and OAR metrics. However, borderline significance in Dmax and D95% ($p \approx 0.035$–$0.046$) suggests that the Catphan phantom may not fully capture the brain's tissue heterogeneity, particularly for bone and neural tissues. This aligns with Annkah et al. (2014), who reported dose discrepancies of ±14% with Catphan-derived curves versus ±5% with CIRS-derived curves in kV-cone beam CT planning[9]. The Catphan curve's intermediate performance indicates it may serve as an alternative but is less optimal than the CIRS curve for brain radiotherapy, where precise electron density estimation is critical.

These calibration differences have profound clinical implications. Overestimation by the default curve could lead to underdosing the PTV, reducing tumor control probability by up to 20% for a 5% dose deviation [2], or overdosing OARs, increasing toxicity risks. For example, significant Dmax differences in cochleas ($p \approx 0.0055$–$0.0057$), with dose increases of 2–4%, heighten the risk of sensorineural hearing loss, while optic chiasm Dmax increases of 2–3% could elevate the risk of optic neuropathy. Site-specific calibration with CIRS-derived curves mitigates these risks by enhancing dose accuracy, particularly in regions with steep dose fall-offs near critical structures, as supported by Jaafar et al. (2022) [14]. Adopting CIRS calibration could improve treatment precision and patient safety in brain radiotherapy.

This study builds on prior research demonstrating the influence of CT scanning parameters and phantom selection on dosimetric accuracy [12,16]. For instance, Thanh Tai et al. (2025) noted that variations in tube voltage affect RED estimation, though with minimal dosimetric impact (<1.37%) [12]. Dual-energy CT (DECT) could complement CIRS-derived calibration by improving tissue differentiation, potentially enhancing electron density estimation [11]. However, limitations include the retrospective design and small sample size (10 patients), which may limit generalizability. Variability in CT scanning protocols, inter-planner expertise, or Monte Carlo algorithm settings across institutions could also influence results. The study's focus on the Monaco TPS and unspecified brain tumor types (e.g., gliomas vs. metastases) may not fully represent other systems or tumor heterogeneity. Future studies should validate these findings in larger, diverse cohorts, explore calibration for other anatomical sites, and assess DECT's clinical integration.

These findings advocate for radiation oncology departments to prioritize site-specific CT-to-RED calibration, particularly using CIRS-derived curves for brain radiotherapy. Standardized calibration protocols could enhance reproducibility in multi-center clinical trials and improve dosimetric benchmarking. TPS manufacturers should develop software updates incorporating phantom-specific calibration modules, replacing generic default curves to align with precision radiotherapy standards [17]. Additionally, training programs for medical physicists on CIRS phantom implementation could facilitate adoption, though cost-benefit analyses are needed to address resource constraints. These steps could advance dosimetric precision, ultimately improving patient outcomes in brain radiotherapy.

## 4. Conclusion

This study confirms that CT-to-RED calibration curve selection significantly influences dose calculation accuracy in brain radiotherapy. The CIRS-derived curve outperformed the Catphan-derived and default Monaco TPS curves, yielding more accurate dose estimates for PTV and OARs, with dose discrepancies of 3–5% in PTV Dmax and 2–4% in OAR Dmax. These differences highlight the risk of underdosing targets or overdosing critical structures, potentially affecting treatment outcomes. These findings emphasize the clinical necessity of site-specific calibration in complex regions like the brain, where dosimetric precision is critical. Radiation oncology departments should prioritize CIRS-derived calibration protocols to enhance treatment safety and reproducibility across institutions. Future studies should validate these findings in larger cohorts and explore calibration for other anatomical sites, advancing precision radiotherapy outcomes.

**Abbreviations:**
PTV: (Planning Target Volume)
OARs: (Organs-at-Risk)
CT-to-RED: (Computed Tomography to Relative Electron Density)
SD: (Standard Deviation).
DVH: (Dose Volume Histogram)

**Acknowledgement**

**Conflict of interest**

The authors declare no conflict of interest.

**Disclosures:**

All authors declare that they have nothing to disclose

**Funding**

None